\documentclass[useAMS,usenatbib,usegraphicx]{mn2e}

\newcommand{\xmm}{{\it XMM-Newton}}
\newcommand{\sss}{RX J0513.9-6951}
\newcommand{\rxj}{RXJ0513}

\title[XMM Spectroscopy of RX J0513.9-6951]{XMM Spectroscopy of the Transient 
Supersoft Source RX~J0513.9-6951: probing the dynamic white dwarf photosphere}
\author[K. E. McGowan et al.]{K. E. McGowan$^{1,2}$\thanks{E-mail: km2@mssl.ucl.ac.uk (KEM); pac@saao.ac.za (PAC)}, P. A. Charles$^{3,4}$, A. J. Blustin$^{1}$, M. Livio$^{5}$,
\newauthor
D. O'Donoghue$^{3}$, B. Heathcote$^{6}$ \\
$^{1}$Mullard Space Science Laboratory, Holmbury St. Mary, Dorking, Surrey, RH5 6NT \\
$^{2}$Los Alamos National Laboratory, Los Alamos, NM 87545, USA \\
$^{3}$South African Astronomical Observatory, P.O. Box 9, Observatory, 7935, South Africa \\
$^{4}$School of Physics and Astronomy, University of Southampton, Southampton SO17 1BJ \\
$^{5}$STScI, 3700 San Martin Drive, Baltimore, MD 21218, USA \\
$^{6}$Barfold Observatory, Glenhope, Victoria, 3444, Australia}

\begin{document}

\date{}

\pagerange{\pageref{firstpage}--\pageref{lastpage}} \pubyear{2005}

\maketitle

\label{firstpage}

\begin{abstract}
The highly luminous ($> 10^{37}$ erg s$^{-1}$) supersoft X-ray sources (SSS) 
are believed to be Eddington limited accreting white dwarfs undergoing
surface hydrogen burning.  The current paradigm for SSS involves thermally
unstable mass transfer from a $1-2 \rm {M}_{\sun}$ companion.  However this 
model has never been directly confirmed and yet is crucial for the evolution of
cataclysmic variables in general, and for the establishment of SSS as
progenitors of type Ia supernovae in particular.  The key SSS is
RX J0513.9-6951 which has recurrent X-ray outbursts every $100-200$ d
(lasting for $\sim 40$ d) during which the optical declines by 1 mag.  We
present the first XMM-Newton observations of RX J0513.9-6951 through one of its
optical low states.  Our results show that as the optical low state
progresses the temperature and the X-ray luminosity decrease, behaviour
that is anti-correlated with the optical and UV emission. We find that as the 
optical (and UV) intensity recover the radius implied by the spectral fits 
increases.  The high resolution spectra show evidence of deep absorption 
features which vary during the optical low state.  Our results are consistent 
with the predictions of the white dwarf photospheric contraction model 
proposed by \citet{sou96}.
\end{abstract} 

       
\begin{figure*}
 \includegraphics[width=110mm]{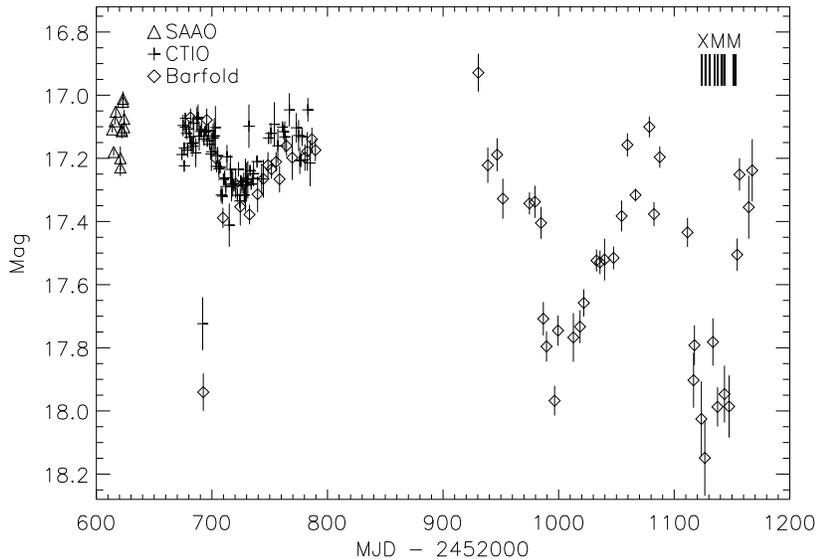}
 \caption{Optical light curve of \sss.  The observations were obtained at 
SAAO (triangles), CTIO (crosses) and Barfold Observatory (diamonds).  The 
times of the start of each X-ray observation are marked.}
 \label{opt_lc}
\end{figure*}

\section{Introduction}

The supersoft X-ray sources (SSS) are a class of X-ray objects
discovered by the {\it Einstein} X-ray Observatory in the late 1970s
\citep{lhg81}.  During the 1990s the class of SSS has been
considerably enlarged through the {\it ROSAT} All-Sky Survey
\citep{tru92}.  Archetypal SSS sources include CAL~83 and CAL~87,
whose defining characteristic is their extremely luminous emission (up
to $L_{\rm bol} \sim 10^{38}$ erg\,s$^{-1}$) at very soft ($<0.5$ keV)
X-ray energies. As such, the SSS are rendered undetectable in the
Galactic plane due to the high level of X-ray absorption, with known
sources lying predominantly in the Magellanic Clouds and M31 
\citep[see e.g.][]{gre96}.

That the SSS are low mass X-ray binaries (LMXBs -
e.g. \citealt{vpm95}) was suggested soon after their discovery by
their optical spectra.  Among the {\it ROSAT} discoveries are several
systems which also show these hallmarks, including \sss, which is the
object of this paper \citep[][hereafter S96]{cow93,sou96}, and for convenience
will be hereafter referred to as \rxj.  The SSS exhibit strong
emission lines of H, He{\sc ii} and higher ionization species, which
probably arise in a hot accretion disk or an associated wind.

Optical spectroscopic and photometric studies \citep[e.g.][]{cal89,sma88} 
yielded some binary information, mainly orbital periods (in the range 
$\sim 10-24$~hrs), yet the nature of the compact object was not clear. 
``Normal'' LMXBs containing neutron star or black hole accretors radiate in 
a wider energy range at the typical luminosities observed in the 
SSS. Using the observed SSS temperatures and bolometric luminosities, van 
den Heuvel et al.\ (1992, hereafter vdH92) therefore proposed that these 
sources in fact contain white dwarf compact objects. The near-Eddington
luminosities are achieved because the white dwarf accretor is able to
sustain steady nuclear burning at its surface. This requires anomalously 
high accretion rates (when compared, for example, to most cataclysmic 
variables; CVs) of $\ga 10^{-7} {\rm M}_{\sun}$~yr$^{-1}$.  However, such 
rates {\it are} attainable if the donor star is comparable to or more 
massive than the white dwarf, so that thermally unstable mass transfer 
occurs \citep[e.g.][]{pac71}.

The vdH92 accreting white dwarf model for the SSS still lacks
convincing observational confirmation.  However, circumstantial
evidence in its favour was presented by S96 who showed that
the observed bipolar outflow emission lines \citep{pak93,cow96} in the
transient SSS, \rxj, have velocities comparable to the escape velocity
of a massive white dwarf.

\rxj\ was discovered in the {\it ROSAT} All Sky Survey \citep{sch93},
and the extended monitoring of the LMC showed its unique
characteristic, namely a very large variability which made it a type of
transient.  The observations showed that the source
brightness increased by a factor of $\sim 20$ during a period of 10
days.  \citet{sch93} found that the PSPC spectrum was well-fitted by a
blackbody of temperature $40$ eV, with a column density of $N_{H} =
0.94 \times 10^{21}$ cm$^{-2}$ and bolometric X-ray luminosity of $\sim 2
\times 10^{38}$ ergs s$^{-1}$.

A key breakthrough in our understanding of the SSS came about through
the MACHO project \citep{alc95} survey of a field in the LMC that
contained \rxj.  This led to the production by S96 of an
exceptional 3.5 year lightcurve of the $V\sim 17$ optical counterpart
which revealed recurrent low states (dropping by $\sim 1$ mag) at
quasi-regular intervals (every $100-200$ days) and remaining low for
$\sim 20-40$ days.  More importantly, the {\it ROSAT} X-ray detections
are {\it only} reported during such (optical) low states, while no
outbursts have been observed during the extensive optical high states.
This showed that the X-ray and optical states are {\it
anti}-correlated.  Furthermore, analysis of the high state optical
data revealed a low amplitude modulation at a period of 18.3 hrs which
S96 interpret as the orbital period of \rxj.

The remarkable behaviour observed in \rxj\ is very difficult to
reconcile with a regular thermonuclear flash model which is normally
accompanied by radius expansion and an increased optical luminosity.
Instead S96 proposed a contraction of the white dwarf from
an expanded, Eddington-limited state to a steady shell-burning phase
as a cause for the X-ray outburst.  This is consistent with {\it all}
the available data, and an X-ray phase due to
contraction of the envelope during shell-burning has been established
observationally for GQ Mus \citep{oge93}.

But what is the cause of such contractions?  S96 proposed that 
this could be due to a decrease in the otherwise very high accretion rate, 
which would also account for the simultaneous optical low states.  This 
then draws an analogy for \rxj\ with the VY Scl class of CVs in that
they experience {\it downward} transitions from an otherwise normal 
{\it on} state \citep{hon04}.  Of course, the shell luminosity will be even 
higher in the optically bright state, but the system is then close to the 
top of the steady burning strip in the $M-M_{WD}$ plane \citep{nom82} which 
means the white dwarf is inflated 
\citep[perhaps by a factor 3 in radius;][]{kov94}.  Hence most of the shell 
luminosity would then be emitted in the UV or EUV.  This leads to the 
prediction that, if this model is correct, as the X-ray outburst evolves it 
will end with its peak emission (or temperature) shifting into the EUV and 
UV.  

We present here the first \xmm\ observations of \rxj\ through one of
its optical low states.  \xmm\ is ideal for this study because, via
its onboard Optical Monitor, we automatically obtain simultaneous UV
photometry.  We use these data to examine the validity of the vdH92
paradigm for SSS via the prediction of S96 for the X-ray
turn-on and turn-off.  If we can confirm directly the vdH92 paradigm
for SSS then this will also establish them as one of the strong
candidates for the progenitors of type Ia supernovae \citep[see
e.g.][]{liv96}.

\section{Observations and Data Reduction}

\subsection{Optical Monitoring}

We were granted ToO status for observations of \rxj\ as part of the
cycle 2 of \xmm.  When the source started its $\sim1$ mag drop in the
optical we would trigger our \xmm\ observations.  To enable this we
set-up a ground based optical monitoring campaign.  From 2002 December
5 (MJD 24520613.5) to 2004 June 11 (MJD 2453167.5) we obtained
$V$-band and unfiltered observations of \rxj\ from SAAO (Sutherland,
South Africa), CTIO (La Serena, Chile), and Barfold Observatory
(Victoria, Australia).  Table~\ref{opt_obs} gives the observing log.
Aperture photometry was performed on all of the observations using
custom packages.  The magnitude of the source was determined using the
magnitude of a comparison star ($V=16.55$; star 2; \citealt{cow93}).
The optical light curve of \rxj\ is shown in Figure~\ref{opt_lc}.

\begin{table*}
\begin{minipage}{145mm}
 \caption{Observation log of the optical measurements of \sss.}
 \label{opt_obs}
 \begin{tabular}{@{}cccrlcccrl}
  \hline
  Date & MJD      & Filter & Exp  & Telescope & Date & MJD      & Filter & Exp  & Telescope \\
       & +2450000 &        & (s)  &           &    & +2450000   &        & (s)  & \\
  \hline
  2002-12-05 & 2613.7858 & $V$ & 30 & SAAO 1.9 m & 
    2003-03-17 & 2715.9919 & $V$ & 180 & CTIO 1.3 m \\  
  2002-12-07 & 2615.0842 & $V$ & 30 & SAAO 1.9 m &
    2003-03-18 & 2716.9884 & $V$ & 180 & CTIO 1.3 m \\   
  2002-12-08 & 2616.0659 & $V$ & 30 & SAAO 1.9 m &
    2003-03-19 & 2717.9898 & $V$ & 180 & CTIO 1.3 m \\ 
  2002-12-08 & 2616.7776 & $V$ & 60 & SAAO 1.9 m &
    2003-03-21 & 2719.0053 & $V$ & 180 & CTIO 1.3 m \\ 
  2002-12-12 & 2620.8108 & $V$ & 30 & SAAO 1.9 m &
    2003-03-22 & 2720.5384 & $none$ & 1800 & Barfold 14 in \\ 
  2002-12-12 & 2620.8119 & $V$ & 60 & SAAO 1.9 m &
    2003-03-23 & 2721.9937 & $V$ & 180 & CTIO 1.3 m \\ 
  2002-12-13 & 2621.9557 & $V$ & 30 & SAAO 1.9 m &
    2003-03-24 & 2722.9926 & $V$ & 180 & CTIO 1.3 m \\ 
  2002-12-13 & 2621.9563 & $V$ & 60 & SAAO 1.9 m &
    2003-03-25 & 2723.9901 & $V$ & 180 & CTIO 1.3 m \\ 
  2002-12-14 & 2622.9846 & $V$ & 30 & SAAO 1.9 m &
    2003-03-26 & 2724.5630 & $none$ & 1260 & Barfold 14 in \\   
  2002-12-14 & 2622.9856 & $V$ & 30 & SAAO 1.9 m &
    2003-03-26 & 2724.9917 & $V$ & 180 & CTIO 1.3 m \\
  2002-12-16 & 2624.0331 & $V$ & 30 & SAAO 1.9 m &
    2003-03-29 & 2727.4295 & $none$ & 1800 & Barfold 14 in \\   
  2002-12-16 & 2624.0343 & $V$ & 30 & SAAO 1.9 m &
    2003-03-27 & 2725.9853 & $V$ & 180 & CTIO 1.3 m \\ 
  2003-02-04 & 2674.0530 & $V$ & 180 & CTIO 1.3 m &
    2003-03-29 & 2727.9840 & $V$ & 180 & CTIO 1.3 m \\
  2003-02-05 & 2675.0242 & $V$ & 180 & CTIO 1.3 m &
    2003-03-30 & 2728.9740 & $V$ & 180 & CTIO 1.3 m \\ 
  2003-02-06 & 2676.0441 & $V$ & 180 & CTIO 1.3 m &
    2003-03-31 & 2729.9751 & $V$ & 180 & CTIO 1.3 m \\
  2003-02-06 & 2676.0674 & $V$ & 180 & CTIO 1.3 m &  
    2003-04-01 & 2730.9747 & $V$ & 180 & CTIO 1.3 m \\ 
  2003-02-07 & 2677.0340 & $V$ & 180 & CTIO 1.3 m &
    2003-04-02 & 2731.9723 & $V$ & 180 & CTIO 1.3 m \\ 
  2003-02-07 & 2677.0832 & $V$ & 180 & CTIO 1.3 m &
    2003-04-03 & 2732.4571 & $none$ & 1710 & Barfold 14 in \\
  2003-02-08 & 2678.0512 & $V$ & 180 & CTIO 1.3 m &   
    2003-04-03 & 2732.9890 & $V$ & 180 & CTIO 1.3 m \\ 
  2003-02-09 & 2679.0976 & $V$ & 180 & CTIO 1.3 m &   
    2003-04-04 & 2733.9926 & $V$ & 180 & CTIO 1.3 m \\ 
  2003-02-10 & 2680.0597 & $V$ & 180 & CTIO 1.3 m &   
    2003-04-05 & 2734.9817 & $V$ & 180 & CTIO 1.3 m \\ 
  2003-02-11 & 2681.0393 & $V$ & 180 & CTIO 1.3 m &   
    2003-04-06 & 2735.9820 & $V$ & 180 & CTIO 1.3 m \\ 
  2003-02-11 & 2681.5979 & $none$ & 1800 & Barfold 14 in &
    2003-04-10 & 2739.0329 & $V$ & 180 & CTIO 1.3 m \\ 
  2003-02-12 & 2682.0321 & $V$ & 180 & CTIO 1.3 m &
    2003-04-10 & 2739.4765 & $none$ & 1440 & Barfold 14 in \\
  2003-02-13 & 2683.0222 & $V$ & 180 & CTIO 1.3 m &
    2003-04-13 & 2742.9679 & $V$ & 180 & CTIO 1.3 m \\ 
  2003-02-14 & 2684.0219 & $V$ & 180 & CTIO 1.3 m &
    2003-04-15 & 2744.4947 & $none$ & 2340 & Barfold 14 in \\
  2003-02-14 & 2684.0281 & $V$ & 180 & CTIO 1.3 m & 
    2003-04-19 & 2748.9734 & $V$ & 180 & CTIO 1.3 m \\ 
  2003-02-16 & 2686.0188 & $V$ & 180 & CTIO 1.3 m & 
    2003-04-19 & 2748.4462 & $none$ & 1710 & Barfold 14 in \\
  2003-02-17 & 2687.0335 & $V$ & 180 & CTIO 1.3 m &
    2003-04-21 & 2750.9637 & $V$ & 180 & CTIO 1.3 m \\ 
  2003-02-18 & 2688.0150 & $V$ & 180 & CTIO 1.3 m &
    2003-04-22 & 2751.4350 & $none$ & 1800 & Barfold 14 in \\
  2003-02-19 & 2689.0135 & $V$ & 180 & CTIO 1.3 m &
    2003-04-24 & 2753.9578 & $V$ & 180 & CTIO 1.3 m \\ 
  2003-02-20 & 2690.0238 & $V$ & 180 & CTIO 1.3 m &
    2003-04-26 & 2755.4181 & $none$ & 1620 & Barfold 14 in \\
  2003-02-21 & 2691.0178 & $V$ & 180 & CTIO 1.3 m &
    2003-04-27 & 2756.9564 & $V$ & 180 & CTIO 1.3 m \\ 
  2003-02-22 & 2692.0747 & $V$ & 180 & CTIO 1.3 m &
    2003-04-29 & 2758.4204 & $none$ & 1800 & Barfold 14 in \\
  2003-02-22 & 2692.6165 & $none$ & 1800 & Barfold 14 in &
    2003-05-01 & 2760.9718 & $V$ & 180 & CTIO 1.3 m \\ 
  2003-02-23 & 2693.0175 & $V$ & 180 & CTIO 1.3 m &
    2003-05-02 & 2761.9844 & $V$ & 180 & CTIO 1.3 m \\ 
  2003-02-24 & 2694.0247 & $V$ & 180 & CTIO 1.3 m &
    2003-05-05 & 2764.4242 & $none$ & 1800 & Barfold 14 in \\
  2003-02-25 & 2695.0212 & $V$ & 180 & CTIO 1.3 m &
    2003-05-03 & 2762.9541 & $V$ & 180 & CTIO 1.3 m \\ 
  2003-02-25 & 2695.4986 & $none$ & 1800 & Barfold 14 in &
    2003-05-07 & 2766.9501 & $V$ & 180 & CTIO 1.3 m \\
  2003-02-26 & 2696.0158 & $V$ & 180 & CTIO 1.3 m &
    2003-05-10 & 2769.4898 & $none$ & 1440 & Barfold 14 in \\
  2003-02-27 & 2697.0121 & $V$ & 180 & CTIO 1.3 m &
    2003-05-13 & 2772.9565 & $V$ & 180 & CTIO 1.3 m \\ 
  2003-02-28 & 2698.0524 & $V$ & 180 & CTIO 1.3 m &
    2003-05-14 & 2773.9495 & $V$ & 180 & CTIO 1.3 m \\ 
  2003-03-01 & 2699.0354 & $V$ & 180 & CTIO 1.3 m &
    2003-05-15 & 2774.9516 & $V$ & 180 & CTIO 1.3 m \\ 
  2003-03-01 & 2699.3983 & $V$ & 180 & CTIO 1.3 m &
    2003-05-16 & 2775.9548 & $V$ & 180 & CTIO 1.3 m \\ 
  2003-03-03 & 2701.0018 & $V$ & 180 & CTIO 1.3 m &
    2003-05-17 & 2776.9572 & $V$ & 180 & CTIO 1.3 m \\ 
  2003-03-04 & 2702.0000 & $V$ & 180 & CTIO 1.3 m &
    2003-05-18 & 2777.9529 & $V$ & 180 & CTIO 1.3 m \\ 
  2003-03-04 & 2702.9992 & $V$ & 180 & CTIO 1.3 m &
    2003-05-21 & 2780.4894 & $none$ & 1980 & Barfold 14 in \\ 
  2003-03-05 & 2703.5417 & $none$ & 1890 & Barfold 14 in &
    2003-05-23 & 2782.4390 & $none$ & 1620 & Barfold 14 in \\
  2003-03-06 & 2704.0338 & $V$ & 180 & CTIO 1.3 m &
    2003-05-23 & 2782.9485 & $V$ & 180 & CTIO 1.3 m \\ 
  2003-03-07 & 2705.0100 & $V$ & 180 & CTIO 1.3 m &
    2003-05-25 & 2784.9439 & $V$ & 180 & CTIO 1.3 m \\ 
  2003-03-08 & 2706.0070 & $V$ & 180 & CTIO 1.3 m &
    2003-05-27 & 2786.5102 & $none$ & 1800 & Barfold 14 in \\
  2003-03-09 & 2707.0093 & $V$ & 180 & CTIO 1.3 m &
    2003-05-30 & 2789.4389 & $none$ & 1800 & Barfold 14 in \\
  2003-03-10 & 2708.0049 & $V$ & 180 & CTIO 1.3 m & 
    2003-10-18 & 2930.5964 & $none$ & 789 & Barfold 14 in \\
  2003-03-11 & 2709.0045 & $V$ & 180 & CTIO 1.3 m &
    2003-10-26 & 2938.6884 & $none$ & 900 & Barfold 14 in \\
  2003-03-11 & 2709.4714 & $none$ & 1890 & Barfold 14 in &
    2003-11-03 & 2946.6754 & $none$ & 900 & Barfold 14 in \\
  2003-03-12 & 2710.0174 & $V$ & 180 & CTIO 1.3 m & 
    2003-11-08 & 2951.6811 & $none$ & 450 & Barfold 14 in \\
  2003-03-13 & 2711.0027 & $V$ & 180 & CTIO 1.3 m &
    2003-12-01 & 2974.6263 & $none$ & 1800 & Barfold 14 in \\
  2003-03-14 & 2712.0351 & $V$ & 180 & CTIO 1.3 m &
    2003-12-06 & 2979.5504 & $none$ & 1710 & Barfold 14 in \\
  2003-03-14 & 2712.9966 & $V$ & 180 & CTIO 1.3 m &
    2003-12-11 & 2984.7006 & $none$ & 135  & Barfold 14 in \\
  2003-03-16 & 2714.9882 & $V$ & 180 & CTIO 1.3 m &
    2003-12-13 & 2986.6378 & $none$ & 900  & Barfold 14 in \\
  \hline
 \end{tabular}
\smallskip

The SAAO observations were taken using the 1.9 m telescope with the WRT1 CCD.  
The CTIO observations were taken using the 1.3 m telescope with the 
ANDICAM-CCD and Fairchild 447 detector.  The Barfold observations were
made using the 14 inch @f/6.0 telescope with the Audine 401E instrument.
\end{minipage}
\end{table*}

\begin{table*}
\begin{minipage}{145mm}
 \contcaption{}
 \begin{tabular}{@{}cccrlcccrl}
  \hline
  Date & MJD      & Filter & Exp  & Telescope & Date & MJD      & Filter & Exp  & Telescope \\
       & +2450000 &        & (s)  &           &   & +2450000    &        & (s)  & \\
  \hline
  2003-12-16 & 2989.5318 & $none$ & 1440 & Barfold 14 in &
    2004-03-23 & 3087.5751 & $none$ & 900 & Barfold 14 in \\
  2003-12-23 & 2996.4976 & $none$ & 1350 & Barfold 14 in &
    2004-04-16 & 3111.4529 & $none$ & 900 & Barfold 14 in \\
  2003-12-26 & 2999.3870 & $none$ & 1350 & Barfold 14 in &
    2004-04-21 & 3116.6381 & $none$ & 1350 & Barfold 14 in \\
  2004-01-08 & 3012.6388 & $none$ & 900 & Barfold 14 in &
    2004-04-22 & 3117.4213 & $none$ & 900 & Barfold 14 in \\
  2004-01-14 & 3018.6219 & $none$ & 900 & Barfold 14 in &
    2004-04-28 & 3123.4436 & $none$ & 900 & Barfold 14 in \\
  2004-01-17 & 3021.6854 & $none$ & 810 & Barfold 14 in &
    2004-05-01 & 3126.4854 & $none$ & 900 & Barfold 14 in \\
  2004-01-28 & 3032.6385 & $none$ & 1350 & Barfold 14 in &
    2004-05-08 & 3133.4803 & $none$ & 900 & Barfold 14 in \\
  2004-01-31 & 3035.6304 & $none$ & 900 & Barfold 14 in &
    2004-05-12 & 3137.3798 & $none$ & 900 & Barfold 14 in \\
  2004-02-04 & 3039.6714 & $none$ & 900 & Barfold 14 in &
    2004-05-18 & 3143.5267 & $none$ & 540 & Barfold 14 in \\
  2004-02-12 & 3047.5078 & $none$ & 720 & Barfold 14 in &
    2004-05-22 & 3147.4069 & $none$ & 900 & Barfold 14 in \\
  2004-02-19 & 3054.5051 & $none$ & 900 & Barfold 14 in &
    2004-05-29 & 3154.4243 & $none$ & 900 & Barfold 14 in \\
  2004-02-24 & 3059.5149 & $none$ & 990 & Barfold 14 in &
    2004-05-31 & 3156.4304 & $none$ & 540 & Barfold 14 in \\
  2004-03-02 & 3066.4980 & $none$ & 900 & Barfold 14 in &
    2004-06-08 & 3164.4391 & $none$ & 900 & Barfold 14 in \\
  2004-03-14 & 3078.4583 & $none$ & 900 & Barfold 14 in &
    2004-06-11 & 3167.5289 & $none$ & 630 & Barfold 14 in \\
  2004-03-18 & 3082.5838 & $none$ & 900 & Barfold 14 in \\
  \hline
 \end{tabular}
\end{minipage}
\end{table*}

On MJD 2452692 \rxj\ faded by $0.6-0.9$ mag, but by the following day
it had re-brightened to its ``steady level''.  A few days later it
experienced another drop in brightness, $\sim 0.25$ mag, which lasted
for $\sim 40$ days.  This less pronounced reduction in brightness can
be seen in earlier lightcurves of \rxj\ (S96), and since the
source did not dim by $>0.5$ mag we did not trigger our \xmm\
over-ride.  A characteristic $1$ mag drop in brightness did
subsequently occur around MJD 2452980, however we were unable to
trigger our observations due to operational restrictions at the \xmm\
Science Operations Centre prevailing at that time.  Fortunately, \rxj\
faded again by $>0.5$ mag on 2004 April 22, and after confirming this
drop over several subsequent nights, we were able to initiate our ToO
on April 26.  The first \xmm\ observation occurred on 2004 April 28,
followed by eight more observations, with the last being taken on 2004
May 28.  The X-ray observation log is given in Table~\ref{xr_obs}.
The times of the start of each X-ray observation are marked in
Figure~\ref{opt_lc}.

\begin{table*}
\begin{minipage}{130mm}
 \caption{Observation log of the X-ray measurements of \sss.}
 \label{xr_obs}
 \begin{tabular}{@{}ccccr}
  \hline
  Observation & Date & MJD       & Instrument & Exp \\
               &      & +2450000 &            & (s)   \\
  \hline
  0151410101 & 2004-04-28 & 3123.7452 & MOS1 & 0     \\
             & 2004-04-28 & 3123.7452 & MOS2 & 0     \\ 
             & 2004-04-28 & 3123.7579 & PN   & 0     \\
  0151412101 & 2004-05-02 & 3127.0789 & MOS1 & 10490 \\
             & 2004-05-02 & 3127.0789 & MOS2 & 10562 \\
             & 2004-05-02 & 3127.0784 & RGS1 & 16885 \\
             & 2004-05-02 & 3127.0784 & RGS2 & 16880 \\ 
             & 2004-05-02 & 3127.0815 & OM $UVW2$ & 2729 \\
             & 2004-05-02 & 3127.1468 & OM $UVW2$ & 2729 \\
  0151412201 & 2004-05-05 & 3130.8146 & MOS1 & 13865 \\
             & 2004-05-05 & 3130.8146 & MOS2 & 14237 \\
             & 2004-05-05 & 3130.8274 & PN & 14008 \\
             & 2004-05-05 & 3130.8141 & RGS1 & 17885 \\
             & 2004-05-05 & 3130.8141 & RGS2 & 17880 \\
             & 2004-05-05 & 3130.8173 & OM $UVW2$ & 5000 \\
             & 2004-05-05 & 3130.9088 & OM $UVW2$ & 3497 \\
  0151412301 & 2004-05-09 & 3134.9117 & MOS1 & 16720 \\
             & 2004-05-09 & 3134.9117 & MOS2 & 16897 \\
             & 2004-05-09 & 3134.9453 & PN & 14049 \\
             & 2004-05-09 & 3134.9111 & RGS1 & 17785 \\
             & 2004-05-09 & 3134.9112 & RGS2 & 17780 \\
             & 2004-05-09 & 3134.9144 & OM $UVW2$ & 5000 \\ 
             & 2004-05-10 & 3135.0267 & OM $UVW2$ & 3397 \\
  0151412401 & 2004-05-12 & 3137.6860 & MOS1 & 6231 \\
             & 2004-05-12 & 3137.6860 & MOS2 & 6443 \\
             & 2004-05-12 & 3137.6987 & PN & 5378 \\
             & 2004-05-12 & 3137.6855 & RGS1 & 25685 \\
             & 2004-05-12 & 3137.6855 & RGS2 & 25680 \\   
             & 2004-05-12 & 3137.6887 & OM $UVW2$ & 3888 \\
             & 2004-05-12 & 3137.7673 & OM $UVW2$ & 3397 \\
             & 2004-05-12 & 3137.8403 & OM $UVW2$ & 3397 \\
             & 2004-05-12 & 3137.9133 & OM $UVW2$ & 3397 \\
  0151412501 & 2004-05-16 & 3141.0642 & MOS1 & 13266 \\
             & 2004-05-16 & 3141.0642 & MOS2 & 13287 \\
             & 2004-05-16 & 3141.0769 & PN & 12192 \\
             & 2004-05-16 & 3141.0636 & RGS1 & 13885 \\
             & 2004-05-16 & 3141.0637 & RGS2 & 13880 \\
             & 2004-05-16 & 3141.0669 & OM $UVW2$ & 5000 \\ 
             & 2004-05-16 & 3141.1584 & OM $UVW2$ & 3097 \\
  0151412601 & 2004-05-18 & 3143.3513 & MOS1 & 11882 \\
             & 2004-05-18 & 3143.3513 & MOS2 & 12043 \\
             & 2004-05-18 & 3143.3641 & PN & 10958 \\
             & 2004-05-18 & 3143.3508 & RGS1 & 13885 \\
             & 2004-05-18 & 3143.3508 & RGS2 & 13880 \\
             & 2004-05-18 & 3143.3540 & OM $UVW2$ & 3819 \\
             & 2004-05-18 & 3143.4319 & OM $UVW2$ & 4278 \\
  0151412701 & 2004-05-26 & 3151.2400 & MOS1 & 17175 \\
             & 2004-05-26 & 3151.2401 & MOS2 & 17304 \\
             & 2004-05-26 & 3151.2528 & PN & 16174 \\
             & 2004-05-26 & 3151.2395 & RGS1 & 18085 \\
             & 2004-05-26 & 3151.2395 & RGS2 & 18080 \\
             & 2004-05-26 & 3151.2427 & OM $UVW2$ & 2819 \\
             & 2004-05-26 & 3151.3090 & OM $UVW2$ & 3278 \\
             & 2004-05-26 & 3151.3806 & OM $UVW2$ & 3278 \\
  0151412801 & 2004-05-28 & 3153.2732 & MOS1 & 15252 \\
             & 2004-05-28 & 3153.2732 & MOS2 & 15278 \\
             & 2004-05-28 & 3153.2860 & PN & 14141 \\
             & 2004-05-28 & 3153.2727 & RGS1 & 15885 \\
             & 2004-05-28 & 3153.2727 & RGS2 & 15880 \\
             & 2004-05-28 & 3153.2759 & OM $UVW2$ & 3094 \\
             & 2004-05-28 & 3153.3454 & OM $UVW2$ & 3094 \\
             & 2004-05-28 & 3153.4149 & OM $UVW2$ & 1000 \\
  \hline
 \end{tabular}
\end{minipage}
\end{table*}

\subsection{XMM-Newton Observations}

The OM \citep{mas01} was operated in Rudi-5 imaging mode during our 
observations using the $UVW2$ filter ($180-225$ nm).  We processed the 
OM images using the \texttt {omichain} routine under the \xmm\ Science 
Analysis System (SAS) v6.0.0.  The photometry was performed with 
\texttt {omsource} using a $5 \arcsec$ radius aperture.  Background 
subtraction was done using counts extracted from an aperture of 
$10 \arcsec$ offset from the source, and deadtime and coincidence loss 
corrections were applied to the count rates.

The EPIC-MOS \citep{tur01} and EPIC-PN \citep{str01} instruments were
operated in timing mode, with the thin filter.  We reduced the EPIC
data with SAS v6.1.0.  The data were filtered to exclude times of high
background.  The first observation (ID 0151410101) was affected by
high radiation and yielded no useful data.  The remaining observations
were filtered on pattern and energy.  Since the data were taken in
timing mode we filtered the data to include only single photon events
for the MOS, and only single and double photon events for the PN.
Only photons with energies in the range $0.3-10$ keV were included.
The data have been filtered to exclude events that may have incorrect
energies, for example those next to the edges of the CCDs and next to
bad pixels.  We note that there are no PN data for the second
observation (ID 0151412101).  The exposure times given in
Table~\ref{xr_obs} reflect the filtering that was performed and
provide the duration of good data.

For each EPIC instrument a spectrum for the source was extracted over
the RAW-X direction, and a region of similar size offset from the
source position was used to extract a corresponding background
spectrum.  We used an extraction region of $>20$ pixels which should 
encompass 100\% of the flux, however, this is uncalibrated.  
In each case we created a photon redistribution matrix
(RMF) and ancillary region file (ARF).  The spectra were regrouped by
requiring at least 30 counts per spectral bin.  The subsequent
spectral fitting and analysis was performed using XSPEC, version
11.3.1.  We extracted source and background light curves from the
filtered event files, which had been barycentrically corrected, using
the same regions.

The RGS \citep{dhe01} spectra were extracted with \texttt {rgsproc} under 
SAS v6.1.0.  Background subtraction was performed with the SAS using regions 
adjacent to that containing the source.  Since SAS v6.1.0 automatically
corrects the response matrices for effective area differences between RGS1 
and RGS2, it was possible to combine the spectra, including both first 
and second order data, channel by channel for each observation using 
the method described by \citet{pag03}.  The resulting combined RGS 
spectrum for each observation was analysed in SPEX 2.00 \citep{kaa02}.

\begin{figure}
 \includegraphics[width=75mm]{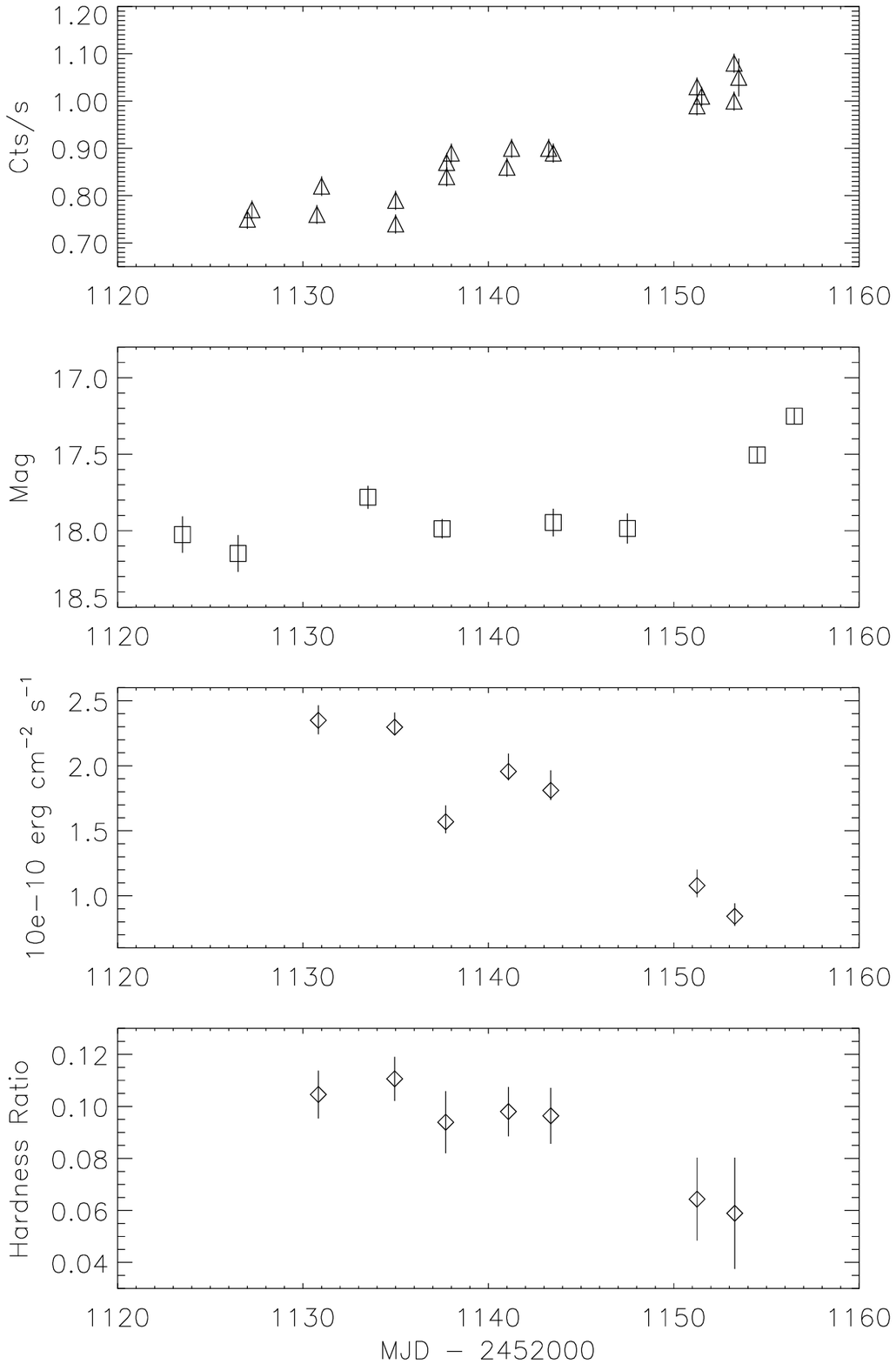}
 \caption{First panel: OM $UVW2$ lightcurve of \sss, second panel: the 
contemporaneous optical data points from Barfold Observatory, third panel: 
EPIC PN lightcurve in the $0.2-10.0$ keV band (see Section 
\ref{sect-lres-spec}), fourth panel: hardness ratio (see Section 
\ref{sect-xr-lc}).}
 \label{uv_lc}
\end{figure}

\section{UV and Optical lightcurves}

We show in Figure~\ref{uv_lc} the OM $UVW2$ lightcurve of \rxj\ with the 
contemporaneous optical data points obtained by Barfold Observatory.  The UV 
intensity is roughly constant at the start of the observations (MJD 
$2453127-2453135$).  By MJD 2453137 the UV flux had started to increase, this 
brightening continued during the remaining observations.  By comparing the 
start of the rise in the UV intensity to the rise in the optical we can see 
that the UV leads the optical.  The lightcurves indicate that the UV increase 
in brightness is more gradual than that in the optical.

We have also plotted the EPIC PN flux in the $0.2-10.0$ keV band (see 
Section \ref{sect-lres-spec}).  The overall trend for the X-rays during our 
observing period is a reduction in the X-ray flux.  There is however a drop 
in X-ray flux at MJD 2453137 which coincides with the start of the rise in the 
UV flux.  The X-ray flux increases after this date, but then continues 
decreasing.  The fourth panel in Figure~\ref{uv_lc} shows the hardness ratio,
which is seen to decrease as the UV and optical intensity increases, see 
Section \ref{sect-xr-lc} for more details.  

\section{X-ray lightcurves and hardness ratios}
\label{sect-xr-lc}

Initial inspection of the X-ray spectra indicates that there is little
emission at energies $>0.8$ keV.  We therefore extracted source and background 
light curves from the PN data in the energy ranges $0.3-0.8$, $0.3-0.5$ and 
$0.5-0.8$ keV.  We binned the lightcurves into 300 s bins.  We define the 
hardness ratio for our data as $(0.5-0.8)/(0.3-0.5)$ keV.  The light curves 
in the different energy bands and the hardness ratios for each observation 
are shown in Figure 3.

The X-ray flux is variable over the course of the observations.  Prominent 
X-ray dips occur in several observations. As the UV and optical brightness 
start to recover ($>$ MJD 2453150) the X-ray flux and hardness ratio 
decrease (see Figure~\ref{uv_lc}, fourth panel).  The dynamical range of the 
variations in the hardness ratio is smaller than the long-term evolution of
the hardness ratio e.g. in the second sequence of data in Figure 3 
(Observation ID 0151412301) two dips occur separated by $\sim 1.5$ hr.

\begin{figure*}
  \vbox to220mm{\vfill 
  \includegraphics[width=91.5mm]{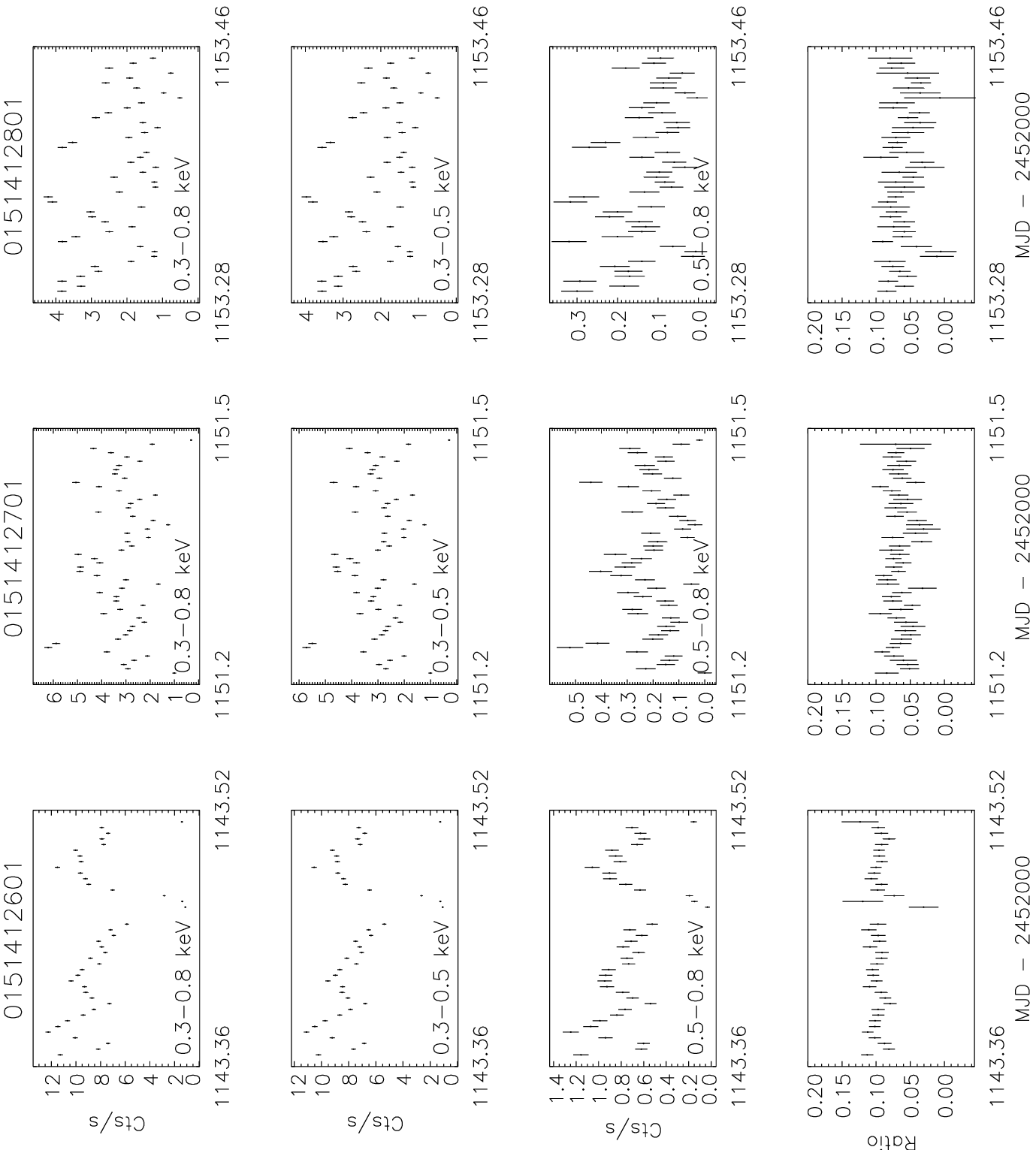}
  \includegraphics[width=91.5mm]{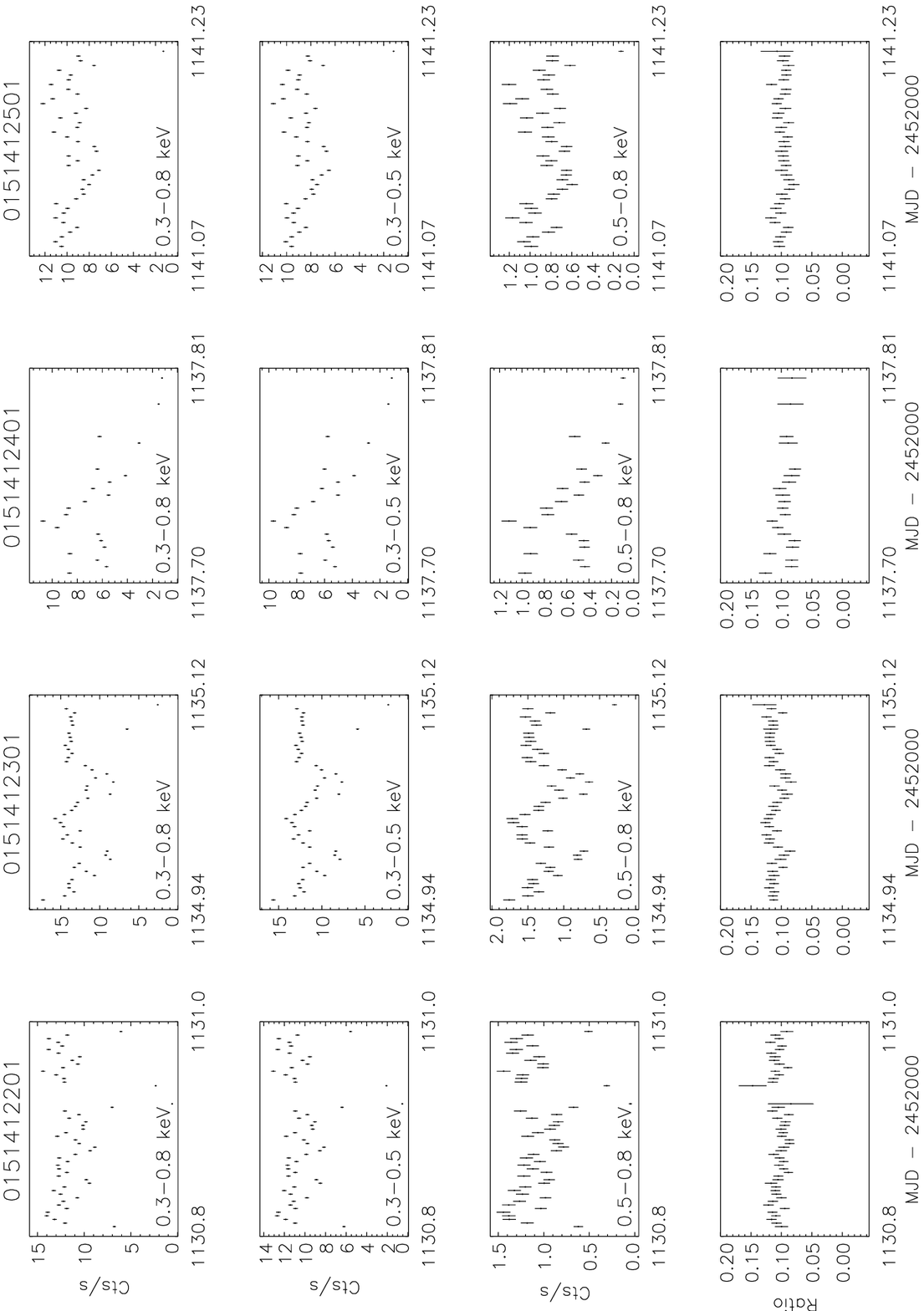}
  \caption{The X-ray light curves for the EPIC PN observations of \sss\ in the 
  $0.3-0.8$ (first panel), $0.3-0.5$ (second panel) and $0.5-0.8$ keV bands 
  (third panel).  The binning for the light curves is 300 s.  The hardness 
  ratio is shown in the fourth panel, where the hardness ratio is defined as 
  $(0.5-0.8)/(0.3-0.5)$.}
 \vfil}
 \label{xr_lc}
\end{figure*}

\section{Low resolution X-ray Spectroscopy}
\label{sect-lres-spec}

For the spectral analysis we have only used the higher signal-to-noise PN 
data.  Since no substantial emission is detected at energies $>0.8$ keV we
extracted the PN spectrum over $0.3-0.8$ keV for each observation.  In each 
case we fitted the PN spectra with a blackbody model modified by neutral 
photoelectric absorption.  

The initial fits to each spectrum where we allowed the equivalent hydrogen 
column density and the temperature to be free resulted in values for $N_{H}$ 
and $kT$ which were poorly constrained, as expected for such low $kT$ values.  
We modelled all the PN spectra simultaneously and found 
$N_{H} = (0.62_{-0.01}^{+0.03}) \times 10^{21}$ cm$^{-2}$ and 
$kT = 43.95_{-3.45}^{+0.55}$ eV.  This value for the equivalent hydrogen 
column density is consistent with the average Galactic absorption in this 
direction\footnote{http://heasarc.gsfc.nasa.gov/ftools/} \citep{bla95}, 
$N_{H} = (0.69 \pm 0.11) \times 10^{21}$ cm$^{-2}$, and the value inferred by 
\citet{gae98} from HST data, $N_{H} = (0.55 \pm 0.10) \times 10^{21}$ 
cm$^{-2}$.  We therefore repeated the fits to the individual PN spectra, 
fixing the column density at the value found from fitting all of the PN 
spectra simultaneously.

\begin{figure}
 \includegraphics[width=75mm]{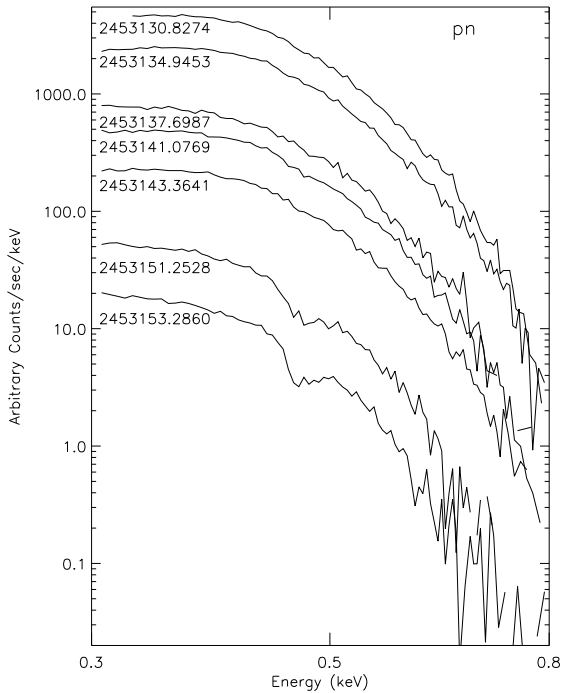}
 \caption{EPIC PN spectra of \sss\ plotted in chronological order of 
descending time.  Each spectrum is labelled with its observation time in MJD.  
At higher energies the spectra are dominated by noise and not by real 
spectral features.}
 \label{epic_spec}
\end{figure}

We find that the blackbody model does not fit the spectra well (typically 
$\chi_{\nu}^{2} > 1.9$), implying that a more sophisticated model is 
required to describe the X-ray emission from \rxj.  At higher energies the
spectra are dominated by noise and not by real spectral features.  We show in 
Figure~\ref{epic_spec} the PN spectra in chronological order.  The last two 
spectra show apparent broad absorption features just below 0.5 keV.  We tried 
adding one or more absorption edges to the fit (at the observed energy they 
could originate from C {\sc v} and/or C {\sc vi}); an edge model was unable 
to reproduce the form of the features. The profile of the absorption is too 
well defined to be explained by an edge, at the spectral resolution of the PN, 
and is thus more indicative of discrete line absorption (see Section 
\ref{sect-rgs}).

\begin{figure}
 \includegraphics[width=86mm]{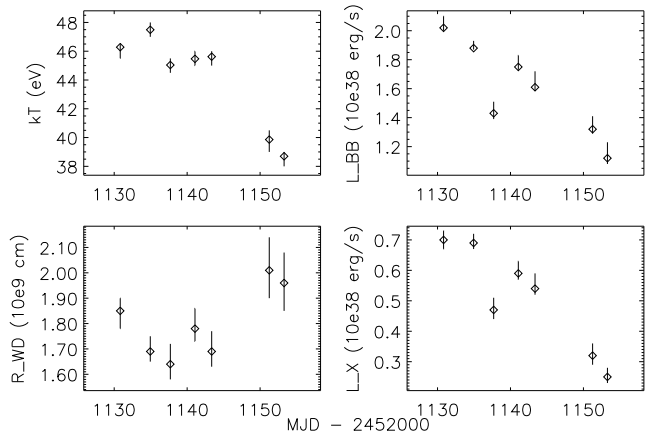}
 \caption{The evolution of $kT$, blackbody luminosity ($L_{BB}$), 
white dwarf radius ($R_{WD}$) and X-ray luminosity in the $0.2-10$ keV band 
($L_{X}$).}
 \label{trend}
\end{figure}

While the blackbody model is a poor fit to the spectra, we can however
investigate the evolution of the spectra by treating the values
obtained as representative of the gross spectral properties.  In
Table~\ref{epic_spec_fit} we show the results of the fits to the PN
spectra.  We find that the temperature varies between $38.7-47.5$ eV, and the 
blackbody luminosity between $(1.1-2.0) \times 10^{38}$ erg s$^{-1}$.
These values are consistent with those found by \citet{sch93} from the 
{\it ROSAT} data, as is the value for $N_{H}$.  We show in Figure~\ref{trend} 
the evolution of $kT$, blackbody luminosity, white dwarf radius and 
X-ray luminosity in the $0.2-10$ keV band, during our observations.

As the source starts to emerge from the optical low state we see that the 
temperature and X-ray luminosity both decrease.  This is anti-correlated with 
the optical and UV emission.

\begin{table*}
\begin{minipage}{180mm}
 \caption{Blackbody fits to the EPIC PN spectra of \sss.}
 \label{epic_spec_fit}
 \begin{tabular}{@{}ccccccccc}
  \hline
  MJD & $kT$ & $L_{BB}$ & $R_{WD}$ & 
  $F_{X}^{a}$ & $F_{X}^{b}$ & $L_{X}^{a}$ & $L_{X}^{b}$ & $\chi_{\nu}^{2}$ \\
  +2450000 & eV & $10^{38}$ erg s$^{-1}$ & 
      $10^{9}$ cm & $10^{-10}$ erg cm$^{-2}$ s$^{-1}$ 
  & $10^{-10}$ erg cm$^{-2}$ s$^{-1}$ & $10^{38}$ erg s$^{-1}$ 
  & $10^{38}$ erg s$^{-1}$ & \\
  \hline
  3130.8274 & $46.28_{-0.78}^{+0.22}$ & $2.02_{-0.02}^{+0.08}$ & $1.85_{-0.07}^{+0.05}$ & $2.35_{-0.11}^{+0.11}$ & $5.33_{-0.11}^{+0.22}$ & $0.70_{-0.03}^{+0.03}$ & $1.60_{-0.03}^{+0.07}$ & 6.6 \\
\\
  3134.9453 & $47.49_{-0.49}^{+0.51}$ & $1.88_{-0.01}^{+0.05}$ & $1.69_{-0.04}^{+0.06}$ & $2.30_{-0.06}^{+0.11}$ & $5.02_{-0.06}^{+0.16}$ & $0.69_{-0.02}^{+0.03}$ & $1.50_{-0.02}^{+0.05}$ & 8.7 \\
\\
  3137.6987 & $45.04_{-0.54}^{+0.46}$ & $1.43_{-0.04}^{+0.08}$ & $1.64_{-0.06}^{+0.08}$ & $1.57_{-0.09}^{+0.13}$ & $3.72_{-0.14}^{+0.23}$ & $0.47_{-0.03}^{+0.04}$ & $1.11_{-0.04}^{+0.07}$ & 1.9 \\
\\
  3141.0769 & $45.47_{-0.47}^{+0.53}$ & $1.75_{-0.02}^{+0.08}$ & $1.78_{-0.05}^{+0.08}$ & $1.96_{-0.07}^{+0.14}$ & $4.57_{-0.09}^{+0.23}$ & $0.59_{-0.02}^{+0.04}$ & $1.37_{-0.03}^{+0.07}$ & 5.8 \\
\\
  3143.3641 & $45.62_{-0.62}^{+0.38}$ & $1.61_{-0.02}^{+0.11}$ & $1.69_{-0.06}^{+0.08}$ & $1.81_{-0.07}^{+0.15}$ & $4.20_{-0.08}^{+0.30}$ & $0.54_{-0.02}^{+0.05}$ & $1.26_{-0.02}^{+0.09}$ & 4.1 \\
\\
  3151.2528 & $39.86_{-0.86}^{+0.64}$ & $1.32_{-0.03}^{+0.09}$ & $2.01_{-0.11}^{+0.13}$ & $1.08_{-0.09}^{+0.12}$ & $3.17_{-0.13}^{+0.26}$ & $0.32_{-0.02}^{+0.03}$ & $0.95_{-0.04}^{+0.08}$ & 5.2 \\
\\
  3153.2860 & $38.70_{-0.70}^{+0.30}$ & $1.12_{-0.04}^{+0.11}$ & $1.96_{-0.11}^{+0.12}$ & $0.84_{-0.07}^{+0.10}$ & $2.62_{-0.14}^{+0.26}$ & $0.25_{-0.02}^{+0.03}$ & $0.78_{-0.04}^{+0.08}$ & 3.2 \\
  \hline
 \end{tabular}
\medskip

The value of $N_{H}$ is fixed at $0.62 \times 10^{21}$ cm$^{-2}$ for all of 
the fits.  $L_{BB}$ is the blackbody luminosity and $R_{WD}$ is the radius 
of the white dwarf.  $F_{X}^{a}$ and $F_{X}^{b}$ are the X-ray fluxes in the 
(a) $0.2 - 10.0$ keV and (b) $0.1 - 2.4$ keV energy ranges, respectively.  
$L_{X}^{a}$ and $L_{X}^{b}$ are the X-ray luminosities in the 
(a) $0.2 - 10.0$ keV and (b) $0.1 - 2.4$ keV energy ranges, respectively.  The 
errors quoted are at the 90\% confidence level.

\end{minipage}
\end{table*}

\section{High resolution X-ray Spectroscopy}
\label{sect-rgs}

\begin{figure}
 \includegraphics[width=75mm]{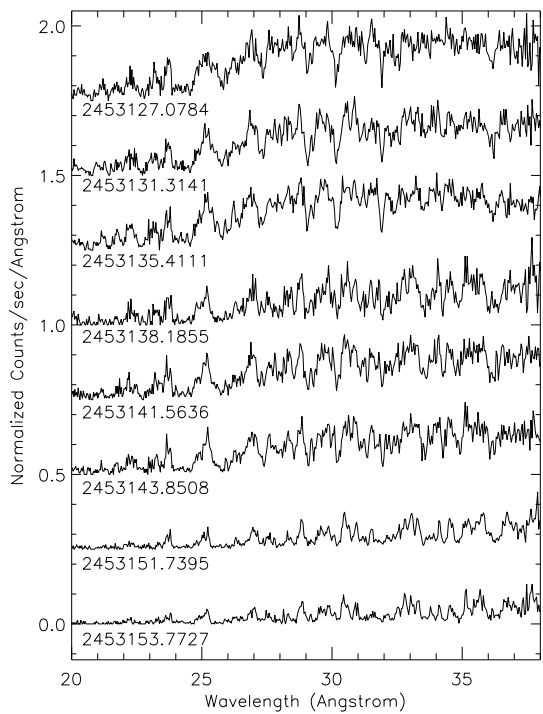}
 \caption{The eight RGS spectra taken during the X-ray high state of \rxj, 
in increasing time order from top to bottom. The spectra have been offset by 
0.25 counts s$^{-1}$ \AA$^{-1}$ in the vertical direction for clarity.}
 \label{all_rgs}
\end{figure}

RGS spectra were available for eight observations.  All of these spectra, 
which are of very high statistical quality, show evidence of deep absorption 
features and probably some low significance narrow emission lines.   A plot of 
all eight spectra is given in Figure~\ref{all_rgs}. 

\begin{figure*}
 \includegraphics[width=115mm]{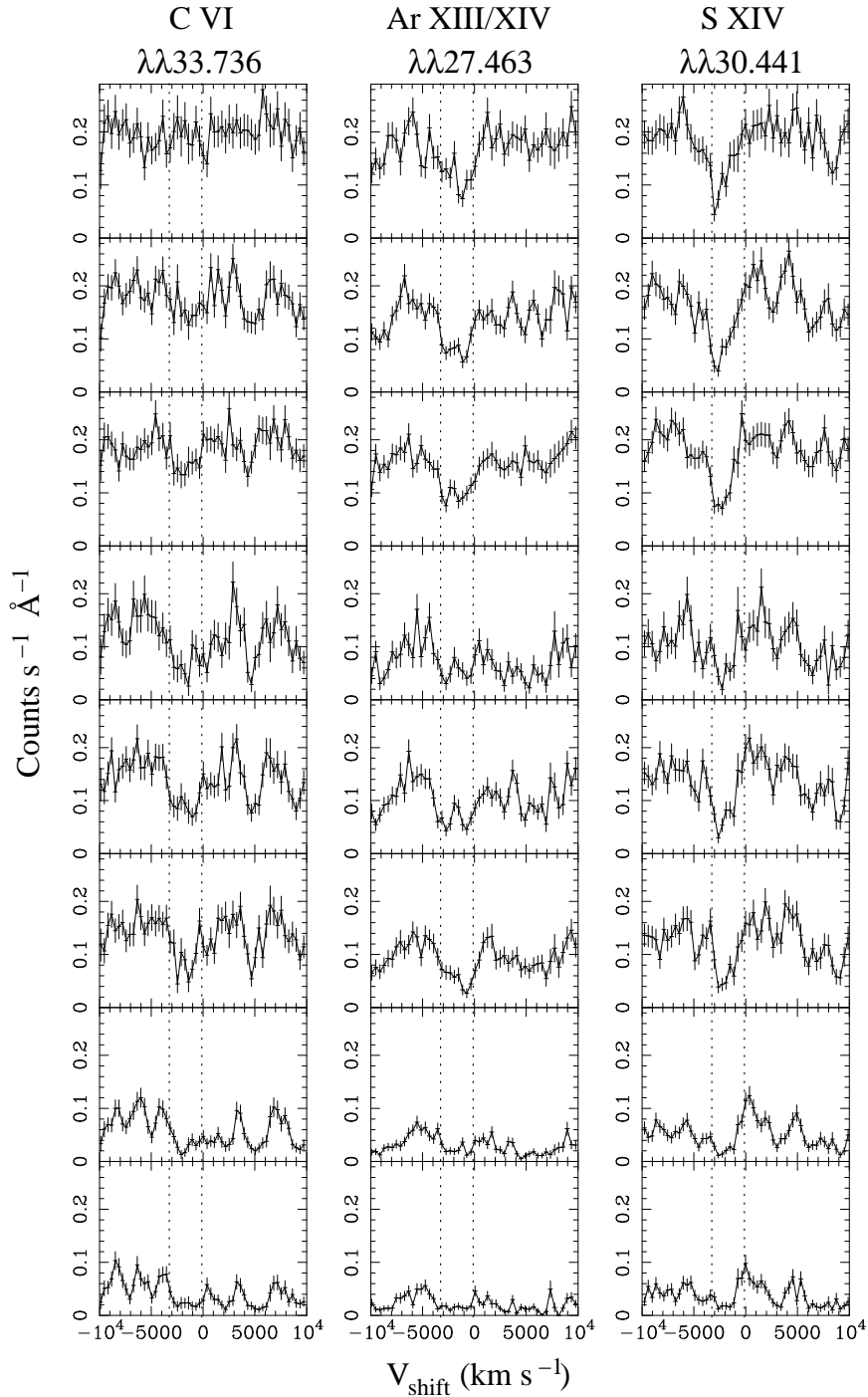}
 \caption{Absorption features from the RGS spectra plotted in velocity space; 
C {\sc vi} Ly$\alpha$, an Ar {\sc xiii}/Ar {\sc xiv} blend and a S {\sc xiv} 
line at 33.736 \AA, 27.463 \AA, and 30.441 \AA, respectively. The 
dotted lines at 0 km s$^{-1}$ and $-3000$ km s$^{-1}$ mark the approximate 
boundaries of the features.}
 \label{rgs_lines}
\end{figure*}

The absorption lines can be identified with transitions from a range of 
high-ionization ions, including C {\sc vi}, S {\sc xii}, S {\sc xiii}, 
S {\sc xiv}, Ar {\sc xiii}, Ar {\sc xiv}, N {\sc vi} and N {\sc vii}, and
possibly O {\sc vii} at 21.6 \AA.  Figure~\ref{rgs_lines} shows three 
individual features plotted in velocity space: C {\sc vi} Ly$\alpha$ at 
33.736 \AA, Ar {\sc xiii}/Ar {\sc xiv} at 27.463 \AA\ (blended at the spectral 
resolution of the RGS), and S {\sc xiv} at 30.441 \AA. It is clear from this 
plot that the lines are blue-shifted and that there are two or more velocity 
components; the outflow velocity along our line of sight is anywhere between 
zero and $\sim 3000$~km s$^{-1}$. 

Figure~\ref{rgs_lines} also indicates that the optical depth of the absorbers 
increases over the course of the eight observations. This is seen most 
clearly in the case of C {\sc vi} Ly$\alpha$, which is barely present in the 
first spectrum, deepens significantly and becomes heavily saturated by the 
time that the final data were taken. The velocity structure of the medium is 
best indicated by the S {\sc xiv} and Ar {\sc xiii}/Ar {\sc xiv} features; 
although there are some apparent changes over the course of the observations, 
there appear to be two main velocity components at $\sim -1000$~km s$^{-1}$ 
and $\sim -3000$~km s$^{-1}$ respectively. The outflow velocities have not
been corrected for the systematic velocity of the LMC of 280 km s$^{-1}$ 
\citep[c.f.][]{hut02}.

\begin{figure*}
 \includegraphics[width=12.7cm]{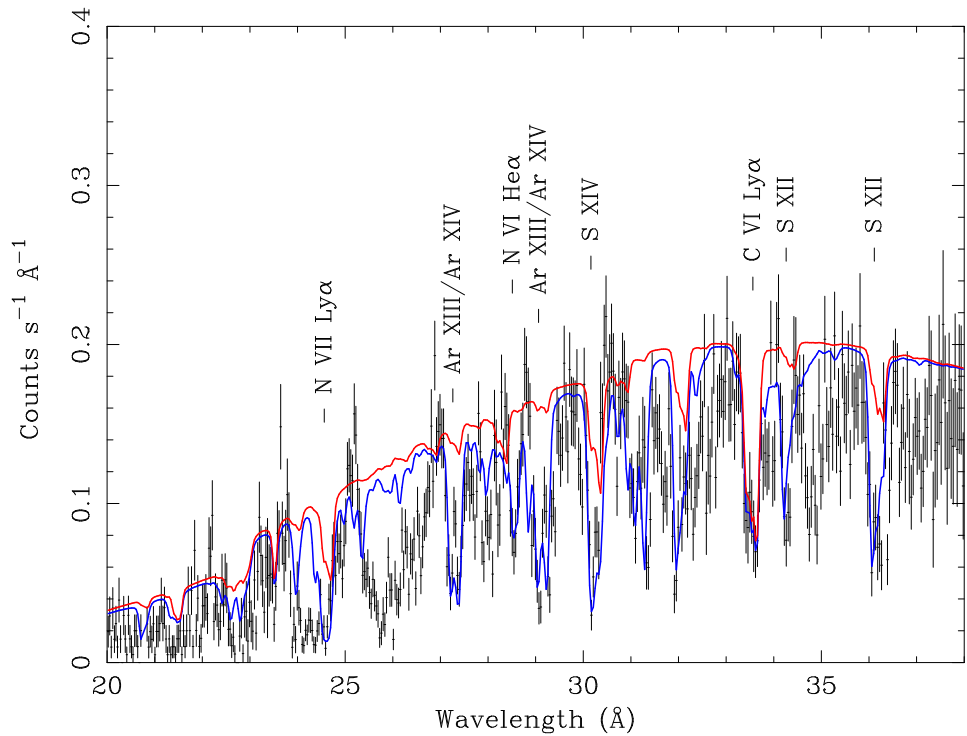}
 \caption{RGS spectrum number 5 (MJD 2453141.5636) compared with a simple 
photoionized absorption model. The dotted line (online version: red) shows a 
version of the model with \citet{and89} Solar abundances, whilst the solid 
line (online version: blue) is a model with enhanced abundances of Ar, N and S 
with respect to Carbon. The positions of various important features are 
labelled.}
 \label{rgs_spec}
\end{figure*}

Figure~\ref{rgs_spec} shows a comparison of the fifth RGS spectrum (MJD 
2453141.5636) with a simple photoionized absorption model to highlight the 
positions of the major features. The model continuum was chosen to represent 
the overall shape of the first RGS spectrum (MJD 2453127.0784), which has the 
least amount of intrinsic ionized absorption and therefore the most bare 
continuum visible. The model continuum consists of a Compton-scattered 
blackbody at a temperature of 70~eV, with Galactic neutral absorption at the 
average column derived from the fits to the PN data ($0.62 \times 10^{21}$ 
cm$^{-2}$). The intrinsic ionized absorption is represented using the 
\emph{xabs} photoionized absorption model in SPEX 2.00 \citep{kaa02}, which 
is based on Xstar \citep{kal82} output and applies absorption by a column of 
gas at a given ionization parameter $\xi$ and specified (variable) elemental 
abundances. The ionization parameter is defined here as :
   \begin{equation}
      \xi = \frac{L}{n r^{\mathrm{2}}} 
   \end{equation}
where $L$ is the source luminosity (in erg s$^{-1}$) in the 13.6~eV$-$13.6~keV 
range (i.e. the entire photoionizing continuum), $n$ the gas density (in 
cm$^{-3}$) and $r$ the source distance in cm, so $\xi$ has units of erg cm 
s$^{-1}$ \citep{tar69}.

The columns of the absorption phases were estimated by matching the depth of 
the modelled C {\sc vi} Ly$\alpha$ feature at 33.736 \AA\ (rest wavelength) 
with the corresponding feature in the data, assuming that carbon is at a 
Solar abundance in the absorbing gas.  A gas turbulent velocity of 200~km 
s$^{-1}$ was chosen (which is below the velocity resolution of the RGS; the 
velocity width of the lines is difficult to establish due to the presence of 
heavily blended features at a range of outflow velocities).  The dot-dash line 
in Figure~\ref{rgs_spec} (red in online version) shows the model with all 
elements at Solar abundances.  Clearly, the depths of the observed features 
require far higher relative abundances of at least N, S and Ar, and the solid 
line (blue in online version) shows the model with abundances of these 
elements enhanced so that the depths of the relevant features are reproduced.  
The spectrum tails off in the region where oxygen absorption is expected, and 
so it was not possible to gain a useful estimate of the relative abundance of 
oxygen. 

The outflow velocities were set from the observed wavelength distributions of 
the C {\sc vi} Ly$\alpha$ and S {\sc xiv} (30.441 \AA) features. Ionization 
parameters of the three phases were then estimated by matching the patterns 
of S absorption in the model phases to those in the data (note that the 
ionization balance of the model was calculated assuming a soft excess AGN SED, 
which is a rather harder spectrum than is strictly appropriate for this 
source).  Table 4 gives the absorbing columns, ionization 
parameters and abundances relative to Solar of C, N, S and Ar of the final 
model.  We emphasize that this model has not been formally fitted to the 
data, and is presented for comparison purposes.

\begin{table*}
\begin{minipage}{145mm}
 \label{rgs_model}
 \caption{Parameters of the photoionized absorber model compared with \rxj\ in 
Figure~\ref{rgs_spec} (solid line; blue in online version). The dot-dash line 
(online: red) in that Figure has identical parameters except that all 
elemental abundances are Solar. The table lists the following parameters for 
each absorbing Phase: $N_{H}$, the absorbing column in units of 
$10^{21}$~cm$^{-2}$; log $\xi$, the log of the ionization parameter where 
$\xi$ is in erg cm s$^{-1}$; v$_{\rm blue}$, the outflow speed in km~s$^{-1}$; 
$A_{C}$, abundance of carbon as a ratio to its Solar abundance; $A_{N}$, 
abundance of nitrogen as a ratio to its Solar abundance; $A_{S}$, abundance 
of sulphur as a ratio to its Solar abundance; $A_{Ar}$, abundance of argon as 
a ratio to its Solar abundance.}
 \begin{tabular}{@{}cccccccc}
  \hline
  Phase & $N_{H}$ & log $\xi$ & v$_{\rm blue}$ & $A_{C}$ & $A_{N}$ & $N_{S}$ & $N_{Ar}$  \\
   \hline
  1     & 10 & 2.6 & 800  & 1 & 1  & 2  & 80  \\
  2     & 2  & 2.4 & 1800 & 1 & 20 & 4  & 10  \\
  3     & 4  & 2.6 & 2800 & 1 & 80 & 30 & 90  \\
  \hline
 \end{tabular}
\end{minipage}
\end{table*}

We note that the columns required to reproduce the deep absorption features 
are one to two orders of magnitude higher than the neutral absorption column 
fitted to the PN spectra, which was consistent with the Galactic absorption 
in the direction of \rxj. This is unsurprising since the very highly-ionized 
gas giving rise to the intrinsic absorption features requires much higher 
columns than neutral material in order to leave a spectral signature, and, at 
least in the earliest RGS observations, discrete spectral line absorption 
dominates over continuum (edge) absorption. By the time the final observations 
were taken, however, the deep absorption features visible even at the PN
spectral resolution are an indication that the intrinsic ionized absorption 
column has become much deeper. 

It is possible that the apparent highly anomalous abundances of N, S and Ar 
are the result of partial covering; if the covering factors of the phases are 
less than unity, but the major absorption lines are saturated, increasing the 
intrinsic column makes low-abundance ions appear unexpectedly prominent 
\citep[e.g.][]{ara02}.  If the relative elemental abundances in the ionized 
outflows in \rxj\ really are factors of several above Solar values, this may 
imply that the outflows contain the products of hot CNO burning.  As stated 
above, it would be difficult to get an estimate of the oxygen abundance from 
this very soft spectrum due to lack of continuum in the relevant spectral 
region. However, at least some of our very rough estimates of the relative 
abundances of C and N do seem to be in line with the prediction 
\citep[e.g.][]{lan05} that carbon should be highly depleted with respect to 
nitrogen in the atmosphere of a supersoft source. The prediction that the 
He/H abundance ratio should be enhanced in winds or outflows from white 
dwarfs is not possible to test with this dataset since no spectral features 
from these two elements are present in the RGS wavelength range, and in any 
case, they are fully stripped at the ionization levels implied by the 
observed transitions.

Comparing the models with the observed spectrum in Figure~\ref{rgs_spec}, it 
is clear that a simple photoionized absorption model is not an adequate 
representation of the data; this is unsurprising since the assumption of 
absorption from the ground state would not be valid in a high density medium 
such as a white dwarf atmosphere.  It is also very likely that atomic data 
are missing or perhaps inaccurate in the current model for various features 
in the spectrum; identification of ionic species is further complicated by 
the likelihood that the observed features consist of complicated blends of 
many different transitions.  Two of the greatest discrepancies are in the 
ranges $25.5-27$ \AA\ and $31.5-33$ \AA\ which correspond to the positions of 
the C {\sc vi} and C {\sc v} absorption edges respectively. 

Approximating the underlying continuum as a blackbody spectrum is probably 
still too simplistic; the continuum and intrinsic absorption will need to be 
modelled self-consistently in order to reach a more accurate understanding of 
both.  \citet{lan05} have shown that a white dwarf atmosphere model is a good 
representation of the soft X-ray spectrum of the supersoft source CAL 83, 
although they found no evidence for a high-speed outflow in that source.  A 
white dwarf atmosphere model with the addition of winds and outflows may 
indeed be required to reproduce the RGS spectra of \rxj.  Nevertheless, the 
present model demonstrates clearly that much of the soft X-ray spectral 
complexity of \rxj\ is consistent with the presence of highly ionized 
outflowing gas, and detailed modelling is currently underway (Blustin et al., 
in preparation).

It is interesting to ask whether the ionized absorber model can provide any 
insights into the nature of the spectral evolution observed in the soft X-ray 
band as \rxj\ comes out of its optical low state. Further work will be 
required to provide a more certain answer to this, although preliminary 
investigations indicate that much of the variation seen in the RGS spectra 
can be explained by changing absorption columns, with the highest columns in 
the last two spectra.  The ionization levels, range of outflow velocities and 
elemental abundances appear to be broadly similar throughout the series of 
observations.

\section{Discussion}

The white dwarf photosphere contraction model proposed by S96 is
based on the prediction that the rise in the X-rays occurs after the drop
in optical luminosity.  From the vdH92 model, during the optical bright state
the accretion rate is very high ($\ga 10^{-7} \rm {M}_{\sun}$ yr$^{-1}$).  
Under these conditions, the white dwarf is slightly inflated, and a large 
majority of the shell luminosity is likely emitted in the UV.  As the 
accretion rate drops, leading to a reduction in the optical flux, the 
photosphere contracts slightly \citep[e.g.][]{kov94,kat85}, this raises the 
effective temperature and hence produces an increase in the X-ray luminosity.

We do not have X-ray observations before the drop, however, previous 
observations have shown that X-rays are not detected when the source is in 
its optical high state.  Our observations confirm that while \rxj\ is in its 
optical low state we see the source in X-rays.

The behaviour of the UV and optical intensity is roughly correlated in the
optical low state.  However, the UV flux starts to rise before the optical
luminosity increases.  We find that as the optical low state progresses the 
X-ray flux decreases, this is anti-correlated with the optical and UV.  This
implies that as the X-ray outburst is evolving the peak of the emission has
moved into the UV, confirming the prediction of the contraction model.

Although our blackbody fits to the PN spectra are poor, the values implied by 
the fits are indicative of the global evolution of the X-ray emission.  We
find that the temperature and luminosity decrease during the optical low state.
The radius determined from the fits decreases during the first observations.
As the optical (and UV) intensity recover the radius increases.  In the 
context of the contraction model \citep[e.g.][]{rei00,hac03a,hac03b}, the 
enhanced X-ray flux will irradiate the companion star and, either by 
inflating material above the secondary's photosphere and causing it to be 
transferred \citep[e.g.][]{rit88} or by heating the magnetic spot region 
\citep[see][]{par79}, will cause the mass transfer rate to increase again.  
This coincides with the white dwarf photosphere re-inflating, hence the 
radius gets larger, and the X-ray flux decreases again.  Therefore, as \rxj\ 
returns to its optical high state the source is once again not observable in 
X-rays.  

Comparing the temperature and luminosity determined by the blackbody fits 
with the white dwarf tracks in the Hertzsprung-Russel diagram of Iben (1982;
Figure 2) one finds agreement with a massive ($1.3-1.4 M_{\sun}$) white dwarf 
which is somewhat expanded.  For a $1.3 M_{\sun}$ white dwarf the radius at 
the stability line is $\sim 4.5 \times 10^8$ cm.  The radius inferred from the 
PN spectral fits implies a radius expansion by a factor of 3.9.  We note that
\citet{hac03b} derive a larger radius using their model, a factor of $\sim 2$ 
larger than the value we determine from our fits.

Our spectral fits to the PN data indicate that a more complicated model
is needed to describe the X-ray emission.  The higher spectral resolution 
data from the RGS reveal the existence of highly ionized gas outflowing along 
our line of sight at speeds of up to 3000 km s$^{-1}$, with multiple velocity 
components that evolve over time. The optical depth of the ionized absorption 
appears to increase over the course of the eight RGS observations; this is 
borne out by the lower-resolution PN spectra, in which deep absorption 
features become apparent in the final two spectra. The presence of this 
absorption partly explains why the blackbody fits are so poor.

There are a variety of possible causes for the increase in optical depth of
the intrinsic absorption.  We could be viewing the system close to pole-on
(i.e. low inclination) through the (presumably) bipolar outflow, and the 
intrinsic covering factor of the outflow is changing.  Possible explanations
for this include the effects of a magnetic cycle of the optical star 
\citep[c.f. CAL83 and VY Scl stars,][respectively]{alc97,hon04}, the 
expansion of the white dwarf or the system precessing with respect to our 
line of sight \citep{sou97}.  We note that \citet{cow02} found a 83.2 d long 
period in the optical high-state MACHO data of \rxj\ which they attribute to 
accretion disk precession.  Such a period would be shorter than the optical 
high/low state cycle length.

Alternatively, it might not be the bipolar outflow we are seeing, but actually
the process of expansion of the white dwarf surface itself, as highly ionized 
nova by-products stream out from the white dwarf surface and gradually build 
into an optically-thick inflated surface, confined by the gravity of the 
white dwarf.  The increase in optical/UV radiation would then occur as the 
`intrinsic' X-rays are down-scattered in the increasingly optically thick 
outer inflated layer of the white dwarf.  The apparently anomalous abundances 
of Sulphur, Argon and Nitrogen in the ionized outflow may indeed imply, if 
this is not just a covering factor effect due to line saturation 
\citep{ara02}, that the gas originated in the nova nucleosynthesis.  The 
abundances may be modified in the atmosphere of the steady nuclear burning 
white dwarf via hot CNO burning \citep[c.f. CNO abundances for CAL83 using 
NLTE model atmospheres,][]{lan05}.  From the nova evolution model calculations
of \citet{pri95}, it follows that a massive white dwarf accreting at a high 
rate of $10^{-6} M_{\sun}$ yr$^{-1}$ has a recurrence time which can be as 
short as the high/low state cycle length of $100-200$ d of \rxj.

The fact that the observed RGS spectra deviate from the blackbody 
approximation at $24-27$ \AA\ may be due to effects of carbon transitions in 
this range due to a white dwarf atmosphere.  A white dwarf atmosphere with an 
effective temperature as deduced from the blackbody fits to the PN data shows 
that with increasing gravity (white dwarf mass) above the C {\sc vi} edge 
(0.49 keV) the atmospheric density increases.  This shifts the ionization 
equilibrium toward a lower degree of ionization, causing the emission edges 
to turn to absorption \citep{har96,har97}.  For the same gravity the energy 
of the C absorption (edge) shifts to lower energies due to a change of the 
ionization state of C.  Thus the C {\sc v}--{\sc vi} absorption due to the 
white dwarf stellar atmosphere may be at least partly responsible for the 
disagreement found from the blackbody approximation \citep[see][]{pae01,rau05}.

The existence of an ionized absorber in the X-ray spectrum of \rxj\ 
seems plausible as the absorption lines found in the RGS spectrum of the fifth 
(and other) observations of the source require a shift due to a velocity up to 
2800 km s$^{-1}$ which is consistent with the escape velocity of a massive 
white dwarf.  Such absorption lines (but usually with lower line-of-sight 
outflow velocities) are commonly found in the soft X-ray spectra of AGN; 
these were first discovered in the {\it Chandra} LETGS spectrum of the 
Seyfert 1 galaxy NGC 5548 \citep{kaa00} and have since been observed in a 
range of other sources \citep[see~e.g.][]{blu05}.  The soft X-ray absorption 
lines in the spectra of these AGN originate in a low-density photoionized 
medium. In \rxj, however, the existence of absorption structures which cannot 
be explained by a photoionized absorber model (especially those between 
$24-27$ \AA) indicates that the absorption takes place in a very different 
medium, probably at a much higher density.  This would be consistent with an 
origin in a white dwarf atmosphere.

The spectral model we constructed for the fifth RGS spectrum in the series 
does, however, reproduce many spectral features due to the ionized outflow 
in \rxj; in this model, the log ionization parameter $\xi$ of the outflowing 
gas is in the range of 2.4 to 2.6.  The ionization structure for such an 
ionization parameter would be consistent with that of the outflowing version 
of model 8 in \citet{kal82}, which models gas photoionized by a blackbody 
emitter.  However, it remains difficult to obtain consistency between the 
other parameters inferred by the spectral fits to the RGS and EPIC-PN 
spectra, the hydrogen absorbing column density of the ionized gas, the 
bolometric luminosity of the source and the blackbody radius of the source.  
If we write the ionization parameter as $\xi = L / N_{H} R$ where 
$N_{H} = n r^2$, then we obtain with $L = 10^{38}$ erg s$^{-1}$, 
$N_{H} = 10^{22}$ cm$^{-2}$ and log $\xi = 2.6$ a value for 
$R = 2.5 \times 10^{13}$ cm, which is unreasonably large for \rxj\ (the size 
of the radius inferred for the white dwarf in \rxj\ is $2 \times 10^{9}$ cm).  
The discrepancy could be due to the real hydrogen column density being much 
higher than it is in our model.  We estimated our model absorbing columns 
assuming that carbon is at its Solar abundance in the absorber.  If carbon 
is less abundant than this the column density could be underestimated.  The
abundance of carbon in the LMC is only a factor $\sim2-3$ less than the Solar 
abundance which would not have a large effect on the column density.  The SED 
used to calculate the SPEX \emph{xabs} absorber model is harder than the 
apparent underlying continuum of this source, and so the ionization balance 
will not be exactly the same as that of model 8 in \citet{kal82}, and the 
inferred ionization level will also be affected if the density is indeed far 
higher than expected in the AGN-type photoionized medium that the \emph{xabs} 
model assumes (see above).  In addition, the outflowing gas may be ionized by 
just a fraction of the radiation from the nuclear burning white dwarf if it 
is still close to the white dwarf atmosphere.
  
We can estimate the mass outflow rate assuming the accretion is slightly 
super-Eddington.  The Eddington luminosity is 
$L_{Edd} \sim 1.2 \times 10^{38}$ $M / M_{\sun}$ erg s$^{-1}$ 
\citep[e.g.][]{gri02}.  As $\dot{M}_{out} \nu = L_{Edd} / c$ 
\citep[see][]{kin03}, we determine with $M = 1 M_{\sun}$, 
$L_{Edd} = 1.2 \times 10^{38}$ erg s$^{-1}$, and $\nu = 3000$ km s$^{-1}$, 
$\dot{M}_{out} = 2.1 \times 10^{-7} M_{\sun}$ yr$^{-1}$.  This value for the 
mass outflow rate is consistent with the wind mass loss rate estimated by
\citet{hac03b} for \rxj.

We can also directly compare the physical parameters inferred from the 
observational data of the low-state egress of \rxj\ with the parameters 
predicted by the model of \citet{hac03b}, for the high/low state transitions 
of \rxj.  The evolution of the blackbody temperature inferred from the EPIC-PN
data follows roughly the evolution of the temperature as inferred by 
\citet{hac03b}, although the temperatures and the radii inferred from the 
observations are about 25\% higher and a factor of 2 smaller, respectively.  
One prediction of the \citet{hac03b} model are winds from the white dwarf with
a wind mass-loss rate of $\sim 10^{-7} M_{\sun}$ yr$^{-1}$ during this part of
the lightcurve.  We can estimate the optical depth of such a highly ionized 
wind due to Thomson scattering opacity assuming an outflow velocity of 
3000 km s$^{-1}$ and a radius of $10^{9}$ cm.  For a wind mass-loss rate of 
$10^{-7} M_{\sun}$ yr$^{-1}$ we calculate an optical depth of 1 which is 
sufficient to obscure the soft X-ray source during the low-state egress.

\section{Summary}

We have presented a series of eight \xmm\ EPIC-PN and RGS observations during 
the late phase (onset of egress) of an optical low-state of the supersoft 
X-ray source \rxj.  Simultaneous \xmm\ OM and long-term ground based optical 
monitoring are also reported.  We have derived the evolution of the fluxes in
the observed bands and the evolution of the X-ray spectral parameters from a
blackbody spectral fit to the EPIC-PN data.  We find that the temperature and
luminosity decrease, and an indication of an increase in the radius of the 
blackbody emitter with time.  During the late phase of the optical low-state 
we find broad spectral dips in the EPIC-PN spectra just below 0.5 keV.  The 
RGS spectra show deep absorption features (e.g. C {\sc vi}, Ar {\sc xiii/xiv},
S {\sc xiv}) which deepen with time and in addition some weak emission lines.
We model the RGS spectra with a Compton-scattered 70 eV blackbody with 
Galactic neutral absorption and additional intrinsic ionized absorption due to
outflowing gas using a photoionized absorption model.  Our spectral model for 
the fifth RGS spectrum, which was taken in the deepest part of the optical low 
state, requires velocities of the outflowing gas of 800, 1800, and 
2800 km s$^{-1}$, ionization parameters log $\xi$ of 2.4 to 2.6, and hydrogen 
columns of at least $2 \times 10^{21}$ to $10^{22}$ cm$^{-2}$.  We find that 
the spectral model lies significantly above the observed spectrum in the 
$24-24.5$ \AA\ and $25.5-27$ \AA\ regime which is likely to be due to effects 
related to the white dwarf atmosphere.

\section*{Acknowledgments}

This work is based on observations obtained with \xmm, an ESA science
mission with instruments and contributions directly funded by ESA
Member States and NASA.  The authors wish to thank Retha Pretorius and the 
SMARTS consortium observers.  The authors also wish to thank the Fred Jansen
and Norbert Schartel and the XMM-Newton control centre staff for executing
out ToO observations.  KEM would like to thank Gavin Ramsay for advice 
on the data analysis.  AJB would like to thank Ehud Behar and Jelle Kaastra 
for useful advice on atomic data and modelling.  KEM and AJB acknowledge the 
support of the UK Particle Physics and Astronomy Research Council (PPARC).  
Part of this work was supported by NASA grant NNG04EG32I.  We thank the 
referee for helpful and constructive comments.

\label{lastpage}

\end{document}